\documentclass[aps,prd,reprint,nofootinbib,superscriptaddress,floatfix]{revtex4-2}

\usepackage{amsmath,amssymb,amsfonts}
\usepackage{graphicx}
\usepackage{bm}
\usepackage{physics}
\usepackage{hyperref}
\usepackage{color}
\usepackage{orcidlink}
\usepackage{microtype}

\begin{document}

\title{Light sea-quark flavor asymmetry and angular momentum of the nucleon in a scalar-vector spectator model}

\author{Parashmani Thakuria\orcidlink{0000-0002-5652-7835}}
\email{parasht@tezu.ernet.in}
\affiliation{Department of Physics, School of Sciences, Tezpur University, Tezpur, India, Pin-784028}

\author{Madhurjya Lalung\orcidlink{0000-0002-7763-5050}}
\email{mlalung2016@gmail.com}
\affiliation{Department of Physics, Nagaon University, Nagaon, India}

\author{Jayanta Kumar Sarma\orcidlink{0000-0001-8854-0308}}
\email{jks@tezu.ernet.in}
\affiliation{Department of Physics, School of Sciences, Tezpur University,\ Tezpur,\ India,\ Pin-784028\ }

\date{\today}

\begin{abstract}
We present a light-front spectator model that describes the proton as an active sea antiquark paired with a composite scalar-vector spectator. Using a spatial profile based on soft-wall AdS/QCD, we fit our initial parameters to CT18NNLO data and Bacchetta-Radici extractions at an initial scale of $\mu_0^2=1.0~\text{GeV}^2$. By allowing these parameters to evolve dynamically, we extend our distributions up to $\mu^2=100~\text{GeV}^2$. We specifically calculate the sea-quark asymmetry at the SeaQuest scale, predicting a sustained $\bar{d}$ enhancement at high $x$ that is in excellent agreement with recent E906 measurements. Additionally, we calculate the leading chiral-even generalized parton distributions (GPDs) and evaluate the total angular momentum carried by the sea quarks, comparing our findings with established results.
\end{abstract}

\maketitle

\section{Introduction}

Understanding the internal structure of the proton in terms of its fundamental constituents, quarks and gluons, remains a central goal of hadron physics \cite{deur2019spin,boffi2007generalized,kuhn2009spin,filippone2001spin,aidala2013spin,ji1997gauge}. For decades, parton distribution functions (PDFs) have served as our primary tool for this, providing a one-dimensional, probabilistic momentum profile of the partons. However, obtaining a complete three-dimensional spatial and momentum picture requires generalized parton distributions (GPDs). Because GPDs depend on the longitudinal momentum fraction, the momentum transfer squared ($t$), and the skewness ($\xi$), they are accessible only through exclusive processes such as deeply virtual Compton scattering (DVCS) \cite{ji1997deeply,radyushkin1996scaling,belitsky2002theory,Hashamipour:2020kip} and deeply virtual meson production (DVMP) \cite{goloskokov2007longitudinal,goloskokov2008role,goloskokov2010attempt,goloskokov2011transversity}. A wealth of experimental data for these processes has already been collected by facilities like H1 \cite{adloff2002diffractive,aaron2009deeply}, ZEUS \cite{adloff2002diffractive,zeus2009measurement}, HERMES \cite{airapetian2012beam,airapetian2012beam1}, COMPASS \cite{d2004feasibility}, and JLab \cite{stepanyan2001observation,Goharipour:2024atx}.

On the theoretical side, the nonperturbative nature of quantum chromodynamics (QCD) makes calculating GPDs from first principles highly challenging. While Lattice QCD continues to advance rapidly, phenomenological models remain essential for building physical intuition and interpreting experimental data. Over the years, the community has utilized various approaches to model these distributions, including the MIT bag model \cite{ji1997study}, light-front constituent quark models \cite{boffi2007generalized,scopetta2004generalized,choi2002continuity,choi2001skewed}, the Nambu--Jona-Lasinio (NJL) model \cite{mineo2005generalized}, color glass condensate frameworks \cite{goeke2008generalized}, chiral quark-soliton models \cite{goeke2001hard,ossmann2005generalized}, Bethe-Salpeter equations \cite{tiburzi2002exploring,noguera2004generalized}, and meson cloud models \cite{pasquini2006virtual,pasquini2007generalized}.

Parton distributions are formally classified by their ``twist,'' which dictates their kinematic suppression at high energies. Recently, significant theoretical effort has been directed toward understanding higher-twist functions, particularly twist-3 GPDs, because they link directly to quark orbital angular momentum \cite{ji1997gauge,jaffe1990g1,hatta2012twist}, DVCS observables \cite{guo2022twist}, and transverse Lorentz forces \cite{burkardt2013transverse,aslan2019transverse}. These have been explored across several models \cite{mukherjee2002off,mukherjee2003helicity,aslan2020singularities,Sharma:2023ibp,bhattacharya2023chiral,zhang2024twist}. Despite this progress in the higher-twist sector, a complete understanding of fundamental leading-twist phenomena remains elusive. A prominent example is the flavor asymmetry of the light sea quarks, specifically the robust experimental evidence that $\bar{d}(x) > \bar{u}(x)$. Identifying the exact nonperturbative mechanisms driving this asymmetry is critical for understanding how the QCD vacuum fluctuates inside the nucleon.

To address this, we construct an effective light-front model originating from the proton's five-quark $|uudq\bar{q}\rangle$ Fock state~\cite{Choudhary:2023unw}. Rather than solving the highly complex five-body problem, we reduce the dynamics to an effective two-body system. We treat the sea antiquark ($\bar{q}$) as the active parton probed by the virtual photon and group the remaining four quarks (the three valence quarks and the partner sea quark) into a single composite spectator. To preserve the essential spin-flavor symmetry, this spectator is modeled as a coherent superposition of scalar (spin-0) and vector (spin-1) configurations. 

For the spatial light-front wave functions, we adopt functional forms inspired by soft-wall AdS/QCD, which effectively capture the confinement scale. The initial model parameters are fixed by fitting to global unpolarized sea-quark PDF data. However, instead of explicitly integrating the coupled DGLAP equations to reach higher energies, we simulate QCD evolution by allowing the nonperturbative wave-function parameters to run dynamically with the energy scale. Using this parameter-driven evolution, we map the sea-quark asymmetry directly at the SeaQuest~\citep{dove2023measurement} experimental scale of $25.5~\text{GeV}^2$. Finally, we compute the leading chiral-even GPDs at zero skewness. By evaluating the Ji sum rule with these distributions, we extract the total angular momentum carried by the sea antiquarks, offering a clearer perspective on their distinct contribution to the proton spin puzzle.

The remainder of the paper is organized as follows. In Sec.~\ref{Sec:II}, we detail the mathematical framework of the light-front scalar-vector spectator model and discuss the parameter fitting procedure in Sec.~\ref{Sec:III}. Section~\ref{SecIV} outlines our approach to scale evolution and presents the flavor asymmetry results evaluated at the SeaQuest scale. In Sec.~\ref{Sec:V}, we derive the overlap representation for the chiral-even GPDs, present our numerical predictions, and calculate the sea-quark angular momentum via the Ji sum rule. Finally, we summarize our findings and conclude in Sec.~\ref{Conclusion}.

\section{Scalar-Vector Spectator Model for Sea Quarks}
\label{Sec:II}

In this section, we introduce an effective light-front model specifically designed to describe the nonperturbative sea-quark content of the nucleon. Directly solving the full five-quark $|uudq\bar{q}\rangle$ Fock state is computationally and analytically prohibitive. To make the problem tractable while retaining the essential physics, we employ an effective two-body spectator approach. In this framework, we assume that the virtual photon strikes an active sea antiquark ($\bar{q}$), while the remaining four constituent quarks (the three valence quarks and the partner sea quark) are treated collectively as a single composite spectator system.

To respect overall spin-flavor symmetries, this composite spectator system can exist in either a scalar (spin-0, denoted $S$) or a vector (spin-1, denoted $V$) configuration. Consequently, the physical proton state can be expressed as a coherent superposition of these two configurations:
\begin{equation}
|P; \pm\rangle = C_S |\bar{q} S\rangle^\pm + C_V |\bar{q} V\rangle^\pm,
\end{equation}
where $C_S$ and $C_V$ represent the probability amplitudes for the scalar and vector spectator components, respectively. 

We work in a symmetric light-cone frame where the proton has no transverse momentum, yielding $P \equiv \left(P^+, \frac{M^2}{P^+}, \mathbf{0}_\perp\right)$. The active sea antiquark carries a longitudinal momentum fraction $x = p^+ / P^+$ and a transverse momentum $\mathbf{p}_\perp$. By momentum conservation, the composite spectator system carries the remaining momentum, $(1-x)P^+$ and $-\mathbf{p}_\perp$.

For the scalar spectator component, the two-particle Fock-state expansion for a proton with helicity $J = \pm 1/2$ is given by \cite{PhysRevD.22.2157}:
\begin{equation}
\begin{aligned}
|\bar{q} S\rangle^\pm &= \int \frac{dx \, d^2\mathbf{p}_\perp}{2(2\pi)^3 \sqrt{x(1-x)}} \\
&\quad \times \Bigg[ \psi_{+\,s}^{\pm(\bar{q})}(x, \mathbf{p}_\perp) \left| +\frac{1}{2}, s; xP^+, \mathbf{p}_\perp \right\rangle \\
&\quad \quad + \psi_{-\,s}^{\pm(\bar{q})}(x, \mathbf{p}_\perp) \left| -\frac{1}{2}, s; xP^+, \mathbf{p}_\perp \right\rangle \Bigg],
\end{aligned}
\end{equation}
where $s=0$ denotes the helicity of the scalar spectator. 

To construct the light-front wave functions (LFWFs) for this system, we require a parametrization that reliably captures the nonperturbative dynamics. We adopt the mathematical framework originally developed for the light-front quark-diquark model \cite{maji2016light}. While that work successfully mapped the valence structure of the nucleon, its general parametrization, which is modeled after soft-wall AdS/QCD predictions, provides a robust foundation for our sea-antiquark and composite spectator system.  We assume the sea quarks are massless and use a proton mass of $M = 0.938$~GeV.

Adapting their functional forms, the scalar spectator LFWFs are \cite{PhysRevD.22.2157}:
\begin{align}
\psi_{+\,s}^{+(\bar{q})}(x, \mathbf{p}_\perp) &= N_S \varphi_1^{(\bar{q})}(x, \mathbf{p}_\perp), \notag \\
\psi_{-\,s}^{+(\bar{q})}(x, \mathbf{p}_\perp) &= N_S \left( -\frac{p^1 + ip^2}{xM} \right) \varphi_2^{(\bar{q})}(x, \mathbf{p}_\perp), \notag \\
\psi_{+\,s}^{-(\bar{q})}(x, \mathbf{p}_\perp) &= N_S \left( \frac{p^1 - ip^2}{xM} \right) \varphi_2^{(\bar{q})}(x, \mathbf{p}_\perp), \notag \\
\psi_{-\,s}^{-(\bar{q})}(x, \mathbf{p}_\perp) &= N_S \varphi_1^{(\bar{q})}(x, \mathbf{p}_\perp).
\end{align}

Similarly, for the vector spectator component, the state expansion includes the helicity states $\lambda_X = 0, \pm 1$~\cite{Ellis_2009}: 
\begin{align}
|\bar{q} V\rangle^{\pm} &= \int \frac{dx \, d^2\mathbf{p}_{\perp}}{2(2\pi)^3\sqrt{x(1-x)}} \nonumber \\
&\quad \times \Bigg[ \psi^{\pm(\bar{q})}_{++}(x, \mathbf{p}_{\perp}) \left| +\frac{1}{2}, +1; xP^+, \mathbf{p}_{\perp} \right\rangle \nonumber \\
&\qquad + \psi^{\pm(\bar{q})}_{-+}(x, \mathbf{p}_{\perp}) \left| -\frac{1}{2}, +1; xP^+, \mathbf{p}_{\perp} \right\rangle \nonumber \\
&\qquad + \psi^{\pm(\bar{q})}_{+0}(x, \mathbf{p}_{\perp}) \left| +\frac{1}{2}, \ 0; xP^+, \mathbf{p}_{\perp} \right\rangle \nonumber \\
&\qquad + \psi^{\pm(\bar{q})}_{-0}(x, \mathbf{p}_{\perp}) \left| -\frac{1}{2}, \ 0; xP^+, \mathbf{p}_{\perp} \right\rangle \nonumber \\
&\qquad + \psi^{\pm(\bar{q})}_{+-}(x, \mathbf{p}_{\perp}) \left| +\frac{1}{2}, -1; xP^+, \mathbf{p}_{\perp} \right\rangle \nonumber \\
&\qquad + \psi^{\pm(\bar{q})}_{--}(x, \mathbf{p}_{\perp}) \left| -\frac{1}{2}, -1; xP^+, \mathbf{p}_{\perp} \right\rangle \Bigg].
\end{align}

We map our composite spectator directly onto the structural derivation used for the vector diquark \cite{Ellis_2009}. For a proton with $J = +1/2$, the LFWFs are:

\begin{align}
\psi_{+\,+}^{+(\bar{q})}(x, \mathbf{p}_\perp) &= N_1^{(\bar{q})} \sqrt{\frac{2}{3}} \left( \frac{p^1 - ip^2}{xM} \right) \varphi_2^{(\bar{q})}(x, \mathbf{p}_\perp), \notag \\
\psi_{-\,+}^{+(\bar{q})}(x, \mathbf{p}_\perp) &= N_1^{(\bar{q})} \sqrt{\frac{2}{3}} \varphi_1^{(\bar{q})}(x, \mathbf{p}_\perp), \notag \\
\psi_{+\,0}^{+(\bar{q})}(x, \mathbf{p}_\perp) &= -N_0^{(\bar{q})} \sqrt{\frac{1}{3}} \varphi_1^{(\bar{q})}(x, \mathbf{p}_\perp), \notag \\
\psi_{-\,0}^{+(\bar{q})}(x, \mathbf{p}_\perp) &= N_0^{(\bar{q})} \sqrt{\frac{1}{3}} \left( \frac{p^1 + ip^2}{xM} \right) \varphi_2^{(\bar{q})}(x, \mathbf{p}_\perp), \notag \\
\psi_{+\,-}^{+(\bar{q})}(x, \mathbf{p}_\perp) &= 0, \notag \\
\psi_{-\,-}^{+(\bar{q})}(x, \mathbf{p}_\perp) &= 0.
\end{align}

The corresponding LFWFs for the $J = -1/2$ state are:

\begin{align}
\psi_{+\,+}^{-(\bar{q})}(x, \mathbf{p}_\perp) &= 0, \notag \\
\psi_{-\,+}^{-(\bar{q})}(x, \mathbf{p}_\perp) &= 0, \notag \\
\psi_{+\,0}^{-(\bar{q})}(x, \mathbf{p}_\perp) &= N_0^{(\bar{q})} \sqrt{\frac{1}{3}} \left( \frac{p^1 - ip^2}{xM} \right) \varphi_2^{(\bar{q})}(x, \mathbf{p}_\perp), \notag \\
\psi_{-\,0}^{-(\bar{q})}(x, \mathbf{p}_\perp) &= N_0^{(\bar{q})} \sqrt{\frac{1}{3}} \varphi_1^{(\bar{q})}(x, \mathbf{p}_\perp), \notag \\
\psi_{+\,-}^{-(\bar{q})}(x, \mathbf{p}_\perp) &= -N_1^{(\bar{q})} \sqrt{\frac{2}{3}} \varphi_1^{(\bar{q})}(x, \mathbf{p}_\perp), \notag \\
\psi_{-\,-}^{-(\bar{q})}(x, \mathbf{p}_\perp) &= N_1^{(\bar{q})} \sqrt{\frac{2}{3}} \left( \frac{p^1 + ip^2}{xM} \right) \varphi_2^{(\bar{q})}(x, \mathbf{p}_\perp).
\end{align}

Finally, the momentum space wave functions $\varphi_i^{(\bar{q})}(x, \mathbf{p}_\perp)$ are given by the modified soft-wall AdS/QCD prediction~\cite{maji2016light}:

\begin{align}
\varphi_i^{(\bar{q})}(x, \mathbf{p}_\perp) &= \frac{4\pi}{\kappa} \sqrt{\frac{\ln(1/x)}{1-x}} x^{a_i^{\bar{q}}} (1-x)^{b_i^{\bar{q}}}\notag \\
&\hspace{1cm}\times\exp\left[ -\delta^{\bar{q}} \frac{\mathbf{p}_\perp^2}{2\kappa^2} \frac{\ln(1/x)}{(1-x)^2} \right],
\end{align}

where $\kappa = 0.4$~GeV is the standard AdS/QCD scale parameter \cite{chakrabarti2013generalized}. In contrast to valence-quark models, the parameters $a_i^{\bar{q}}$, $b_i^{\bar{q}}$, and $\delta^{\bar{q}}$ in our framework dictate the longitudinal momentum distributions and the transverse momentum profile of the sea antiquarks. We determine these parameters by fitting the model to the unpolarized sea-quark parton distribution functions from the CT18NNLO global analysis~\cite{hou2021new} at the initial scale $\mu_0^2=1.0~\text{GeV}^2$.

\begin{figure*}[htpb]
\centering
\includegraphics[width=0.8\textwidth]{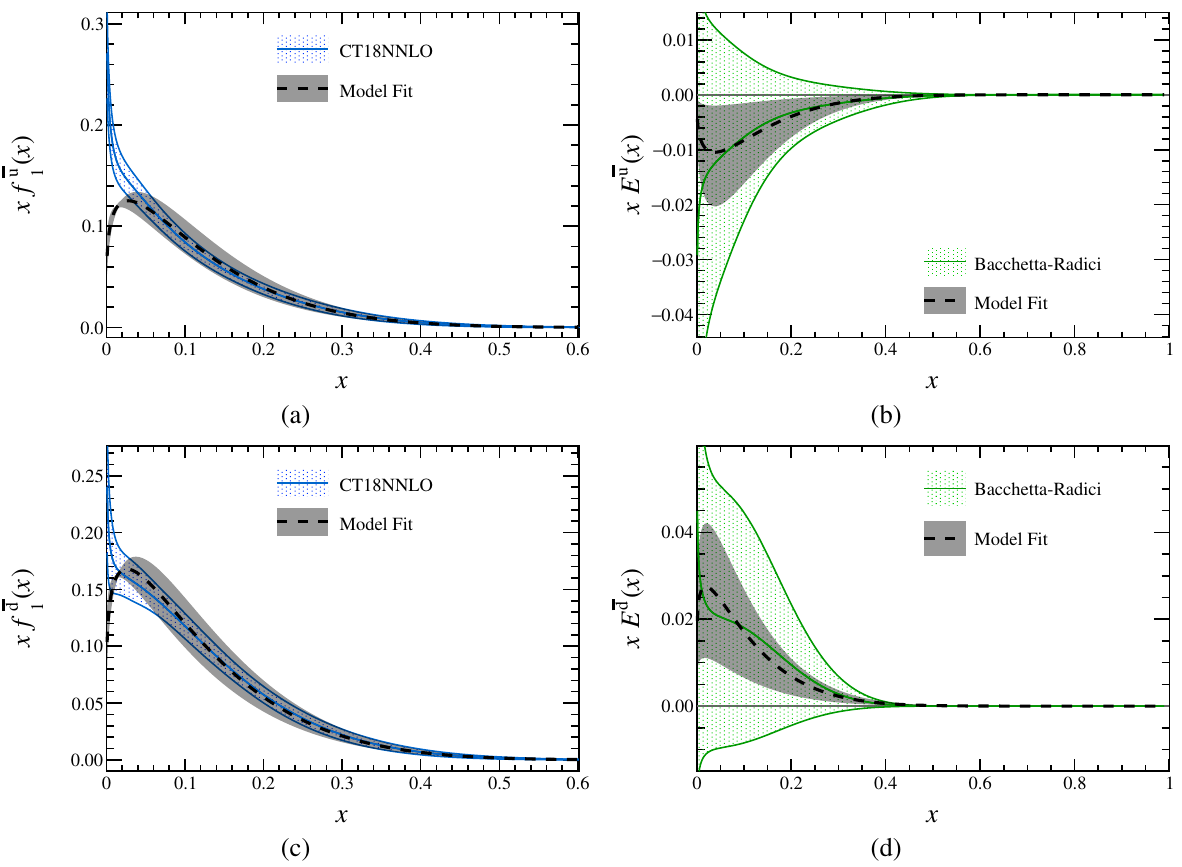}
\caption{The unpolarized sea-quark PDFs in the proton fitted to CT18NNLO data~\cite{hou2021new} at the initial scale $\mu^2_0 = 1.0~\text{GeV}^2$. Panels (a) and (c) display the distributions for $x\bar{u}$ and $x\bar{d}$, respectively, in the kinematic region $0.0001 < x < 1$. Panels \((b)\) and \((d)\) show the corresponding fits for the magnetic distribution $xE(x)$ compared against the Bacchetta-Radici extractions \cite{Bacchetta:2011gx}.}
\label{fit}
\end{figure*}

\begin{table*}[htpb]
\caption{Fitted kinematic model parameters for the $\bar{u}$ and $\bar{d}$ sea quarks at the initial scale $Q_0^2 = 1.0$~GeV$^2$.}
\label{tab:fit_parameters}
\begin{ruledtabular}
\begin{tabular}{ccccc}
Flavor & $a_1^{\bar{q}}$ & $a_2^{\bar{q}}$ & $b_1^{\bar{q}}$ & $b_2^{\bar{q}}$ \\
\colrule
$\bar{u}$ & $-0.381 \pm 0.0563$ & $0.704_{-0.057}^{+0.057}$ & $3.852_{-0.084}^{+0.086}$ & $2.333_{-0.075}^{+0.093}$ \\
$\bar{d}$ & $-0.398 \pm 0.0711$ & $0.568_{-0.025}^{+0.025}$ & $3.430_{-0.051}^{+0.053}$ & $3.342 \pm 0.246$ \\
\end{tabular}
\end{ruledtabular}
\end{table*}

\section{UNPOLARIZED SEA QUARK DISTRIBUTIONS}
\label{Sec:III}
With the light-front wave functions for the sea antiquark and spectator cluster defined, we can now evaluate the unpolarized sea quark parton distribution functions (PDFs). In the light-front formalism, the unpolarized PDF, $f_1(x)$, represents the probability of finding a parton with a longitudinal momentum fraction $x$ inside the nucleon. 

Following the standard overlap representation, the unpolarized PDF is obtained by integrating the square of the LFWFs over the transverse momentum $\mathbf{p}_\perp$ of the active parton. Because our physical state is a superposition of scalar and vector spectator configurations, the total unpolarized sea antiquark distribution $f_1^{\bar{q}}(x)$ is the weighted sum of the contributions from the scalar ($S$) and vector ($V$) components:
\begin{equation}
f_1^{\bar{q}}(x) = C_S^2 f_1^{(S)}(x) + C_V^2 f_1^{(V)}(x).
\end{equation}

The individual contributions are calculated by summing over all possible helicity combinations of the active sea antiquark and the spectator cluster. For the scalar spectator, the unpolarized distribution is:
\begin{equation}
f_1^{(S)}(x) = \int \frac{d^2\mathbf{p}_\perp}{16\pi^3} \left[ \left| \psi_{+\,s}^{+(\bar{q})}(x, \mathbf{p}_\perp) \right|^2 + \left| \psi_{-\,s}^{+(\bar{q})}(x, \mathbf{p}_\perp) \right|^2 \right].
\end{equation}
Similarly, for the vector spectator, we sum over the longitudinal and transverse polarization states of the vector cluster:
\begin{equation}
\begin{split}
f_1^{(V)}(x) &= \int \frac{d^2\mathbf{p}_\perp}{16\pi^3} \left[ \left| \psi_{+\,+}^{+(\bar{q})}(x, \mathbf{p}_\perp) \right|^2 + \left| \psi_{-\,+}^{+(\bar{q})}(x, \mathbf{p}_\perp) \right|^2 \right. \\
&\quad \left. + \left| \psi_{+\,0}^{+(\bar{q})}(x, \mathbf{p}_\perp) \right|^2 + \left| \psi_{-\,0}^{+(\bar{q})}(x, \mathbf{p}_\perp) \right|^2 \right].
\end{split}
\end{equation}

Substituting the AdS/QCD-inspired wave functions from Sec.~\ref{Sec:II} into the overlap integrals and performing the Gaussian integration over transverse momentum yields the analytical expression for the unpolarized sea-antiquark PDF:

\begin{align}
f_{1}^{\bar{q}}(x) &= \left( C_{S}^{2} N_{S}^{2} + C_{V}^{2} \left[ \frac{1}{3} N_{0}^{2} + \frac{2}{3} N_{1}^{2} \right] \right) 
 \bigg[ \frac{1}{\delta^{\bar{q}}} x^{2a_{1}^{\bar{q}}} \notag \\
 &\quad (1-x)^{2b_{1}^{\bar{q}}+1} + \frac{\kappa^{2}}{M^{2} \ln(1/x)} \frac{x^{2a_{2}^{\bar{q}}-2} (1-x)^{2b_{2}^{\bar{q}}+3}}{(\delta^{\bar{q}})^{2}} \bigg].
\label{eq:unpol_sea}
\end{align}

Equation (\ref{eq:unpol_sea}) provides a highly flexible, yet physically grounded, parametrization for the sea quarks. The overall normalization is governed by the $C_{S,V}$ coefficients and the wave function normalizations ($N_S, N_0, N_1$). To extract the specific flavor distributions for the up-sea ($\bar{u}$) and down-sea ($\bar{d}$), one simply adjusts the isospin coefficients ($C_S, C_V$) to project out the appropriate flavor states, and fits the corresponding kinematic parameters ($a_i^{\bar{q}}$, $b_i^{\bar{q}}$, $\delta^{\bar{q}}$) to global dataset extractions at the initial scale $\mu_0$.

\begin{figure*}[htpb]
\centering
\includegraphics[width=1.0\textwidth]{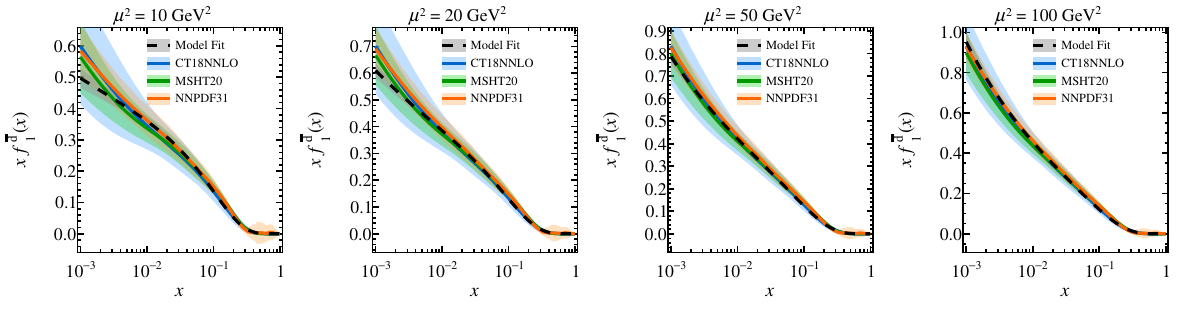}
\caption{Evolution of unpolarized PDF in this model for \(\bar{d}\) is shown at \(\mu^2=10,20,50\) and \(100\,\text{GeV}^2\). Our model predictions are compared with NNPDF31~\cite{ball2017parton}, CT18NNLO~\cite{hou2021new} and  MSHT20 results~\cite{bailey2021parton}.}
\end{figure*}

\begin{figure*}[htpb]
\centering
\includegraphics[width=1.0\textwidth]{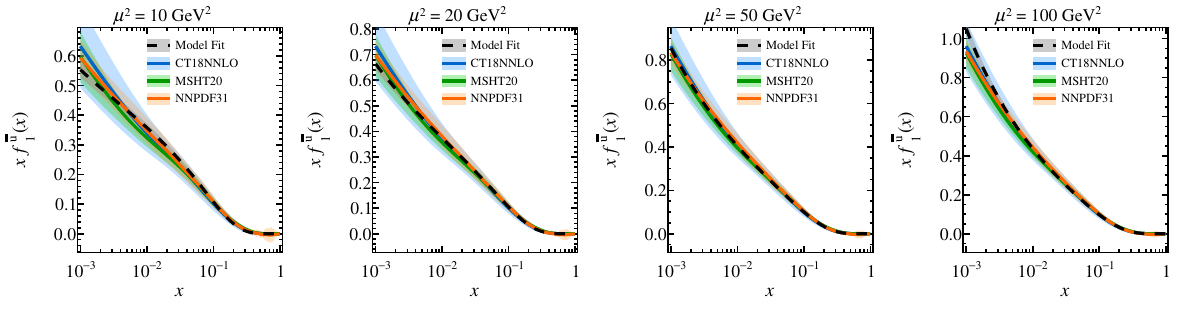}
\caption{Evolution of unpolarized PDF in this model for \(\bar{u}\) is shown at \(\mu^2=10,20,50\) and \(100\,\text{GeV}^2\). Our model predictions are compared with NNPDF31~\cite{ball2017parton}, CT18NNLO~\cite{hou2021new} and  MSHT20 results~\cite{bailey2021parton}.}
\end{figure*}

\subsection{Numerical fitting and model parameters}

There are several parameters ($a_1^{\bar{q}}$, $a_2^{\bar{q}}$, $b_1^{\bar{q}}$, $b_2^{\bar{q}}$, $C_S^2$, and $C_V^2$) in our model for each sea quark flavor that characterize the accuracy of the framework. The parameters $C_S^2$ and $C_V^2$ govern the relative probability amplitudes of the scalar and vector spectator components, respectively. The parameters $a_i^{\bar{q}}$ and $b_i^{\bar{q}}$ dictate the behavior of the distributions in the extreme kinematic limits of $x$; $a_i^{\bar{q}}$ is crucial to the lower-$x$ region, while $b_i^{\bar{q}}$ controls the large-$x$ behavior. For this analysis, we set the AdS/QCD scale parameter to $\kappa = 0.4$~GeV and the transverse momentum drop-off parameter to $\delta^{\bar{q}} = 1.0$. 

We determine these parameters by fitting the model's unpolarized light sea quark distributions to the latest available NNLO data for $x\bar{u}(x)$ and $x\bar{d}(x)$ from the CT18 global analysis~\cite{hou2021new} at the initial scale $Q_0^2 = 1.0$~GeV$^2$. Further constraints are applied through the phenomenological target for the distribution $xE(x)$ derived from the Bacchetta-Radici extractions. This ensures the model accurately captures the spin-orbit correlations alongside the collinear momentum fractions. 

The normalization constants \(N_i^2\) are calculated using the conditions~\cite{PhysRevD.78.074010}:
\begin{equation}
    \int_0^1 f_1^{(S)}(x) \, dx = 1, \quad \text{and} \quad \int_0^1 f_1^{(V)}(x) \, dx = 1.
\end{equation}

We evaluate the fit across 1000 data points distributed logarithmically within the interval $0.001 \le x \le 0.99$ using \texttt{MINUIT}~\cite{james1994minuit}. To stabilize the minimization and regulate the low-$x$ divergence of the model, we impose physical boundary constraints on the parameters, such as $a_1 > -0.5$. Furthermore, to ensure the model preserves the correct longitudinal momentum fractions, we enforce the condition that the truncated second moment of our distribution strictly matches the corresponding integral from the global analysis:
\begin{equation}
    \int_{0.001}^{1} x f_1^{\bar{q}}(x) \, dx = \int_{0.001}^{1} \left[ x \bar{q}(x) \right]_{\text{CT18NNLO}} \, dx.
\end{equation}
Subject to these constraints, we determine the central values of our model parameters by performing a simultaneous global fit. The resulting $\chi^2/\text{d.o.f.}$ for the simultaneous fit of the $\bar{u}$ sea quark distributions is 0.261, whereas for the $\bar{d}$ sea quarks, the $\chi^2/\text{d.o.f.}$ is 0.333. 

To fully capture the parameter correlations and the potentially asymmetric nature of the $\chi^2$ valley at this initial scale, the parameter uncertainties were extracted using the MINOS algorithm within \texttt{MINUIT}~\cite{james1994minuit}. The fitted kinematic parameters $a_i^{\bar{q}}$ and $b_i^{\bar{q}}$, along with their corresponding asymmetric MINOS uncertainties, are listed in Table~\ref{tab:fit_parameters}. The structural coefficients for the $\bar{u}$ sea quark are obtained as $C_S^2 = 0.143 \pm 0.004$ and $C_V^2 = 0.396 \pm 0.010$, which yield the normalization constants $N_S^2 = 0.675$, $N_0^2 = 0.888$, and $N_1^2 = 0.568$. For the $\bar{d}$ sea quark, the coefficients are fitted as $C_S^2 = 0.266 \pm 0.005$ and $C_V^2 = 0.476 \pm 0.009$, resulting in the normalization constants $N_S^2 = 0.550$, $N_0^2 = 0.724$, and $N_1^2 = 0.463$. 

In Fig.~\ref{fit}, we show the results of our fits for the sea quark distributions, $xf_1^{\bar{q}}(x)$ and $xE^{\bar{q}}(x)$, at the scale $Q_0^2 = 1.0$~GeV$^2$. The data bands identify the CT18NNLO~\cite{hou2021new} and Bacchetta-Radici analyzes, while the corresponding lines represent our results computed with the effective parameters.

\section{UNPOLARIZED PDF EVOLUTION}
\label{SecIV}

In perturbative quantum chromodynamics (QCD), the scale evolution of parton distributions is governed by the Dokshitzer-Gribov-Lipatov-Altarelli-Parisi (DGLAP) equations. Unlike valence quarks, which evolve independently as nonsinglet distributions, the evolution of sea antiquarks $\bar{q}(x, \mu^2)$ is intimately coupled to the gluon distribution $g(x, \mu^2)$ through gluon splitting processes. At leading order, the DGLAP equation for the sea antiquark is given by
\begin{align}
\frac{\partial \bar{q}(x, \mu^2)}{\partial \ln \mu^2} &= \frac{\alpha_s(\mu^2)}{2\pi} \int_x^1 \frac{dy}{y} \bigg[ P_{qq}\left(\frac{x}{y}\right) \bar{q}(y, \mu^2) \notag \
&\hspace{3cm}+ P_{qg}\left(\frac{x}{y}\right) g(y, \mu^2) \bigg],
\end{align}
where $P_{qq}$ and $P_{qg}$ are the Altarelli-Parisi splitting functions. The strong coupling constant at one loop is parameterized as
\begin{equation}
\alpha_s(\mu^2) = \frac{4\pi}{\beta_0 \ln(\mu^2/\Lambda_{\text{QCD}}^2)},
\end{equation}
with $\beta_0 = 11 - \frac{2}{3}n_f$ and $\Lambda_{\text{QCD}} = 0.226~\text{GeV}$.

Because our effective light-front spectator model absorbs the multi-parton higher Fock states into an effective two-body system, it does not explicitly track a dynamical gluon distribution $g(x, \mu^2)$. Consequently, directly integrating the coupled integro-differential DGLAP equations within our analytical framework is highly nontrivial. To bypass this limitation while maintaining consistency with QCD evolution, we follow a phenomenological approach~\cite{maji2016light}, in which the scale dependence is absorbed into the nonperturbative kinematic parameters of the light-front wave functions.

We map the unpolarized sea quark distribution $f_1^{\bar{q}}(x, \mu^2)$ at an arbitrary hard scale $\mu^2$ by redefining the static parameters $a_i^{\bar{q}}$, $b_i^{\bar{q}}$, and $\delta^{\bar{q}}$ from Eq.~(\ref{eq:unpol_sea}) as continuous functions of $\mu^2$. Because sea quark distributions evolve rapidly at low $x$ due to the running of the strong coupling, we expand the longitudinal momentum parameters in terms of the logarithmic ratio
\begin{equation}
L = \ln\left[\frac{\alpha_s(\mu_0^2)}{\alpha_s(\mu^2)}\right].
\end{equation}

A purely polynomial expansion in $L$ is unbounded and leads to unphysical parameter growth at large scales. To ensure asymptotic stability, we introduce a bounded rational (Padé-inspired) damping form. The scale-dependent parameters are constructed as
\begin{align}
a_i^{\bar{q}}(\mu^2) &= a_i^{\bar{q}}(\mu_0^2) + c_{a_i}^{\bar{q}} L + d_{a_i}^{\bar{q}} \frac{L^2}{(1+L)^2}, \\
b_i^{\bar{q}}(\mu^2) &= b_i^{\bar{q}}(\mu_0^2) + c_{b_i}^{\bar{q}} L + d_{b_i}^{\bar{q}} \frac{L^2}{(1+L)^2},
\end{align}
for $i \in {1,2}$. This construction retains quadratic flexibility at low scales ($L \to 0$) while smoothly saturating at large $L$, ensuring well-behaved asymptotic evolution.

The transverse momentum width parameter $\delta^{\bar{q}}$ is modeled using a stretched exponential form driven by $L_Q = \ln(\mu^2/\mu_0^2)$ with the same damping structure:
\begin{equation}
\delta^{\bar{q}}(\mu^2) = \exp\left[ \delta_1^{\bar{q}} \left( \frac{L_Q^2}{(1+L_Q)^2} \right)^{\delta_2^{\bar{q}}} \right].
\end{equation}
This ensures $\delta^{\bar{q}}(\mu_0^2)=1$ while preventing unphysical growth at large $\mu^2$.

The ten evolution coefficients ($c_i$, $d_i$, $\delta_{1,2}$) are determined through a global optimization using the \texttt{MINUIT} algorithm. The fit is performed over a grid of 50 $x$ points spanning $x \in [0.001, 0.99]$ and 1500 scale points in the range $\mu^2 \in [1.0, 100.0]$~GeV$^2$, using the CT18NNLO global analysis~\cite{hou2021new} as the reference.

\begin{figure*}[htpb]
\centering
\includegraphics[width=1.0\textwidth]{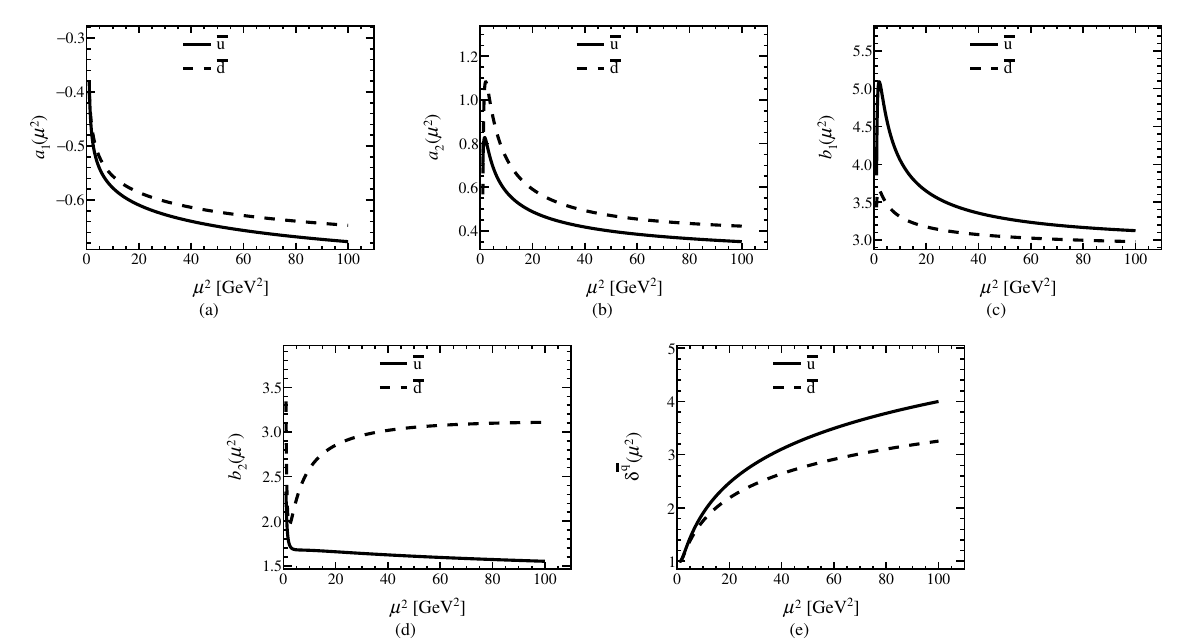}
\caption{Scale dependence of the parameters in the range $\mu^2=1$--$100~\text{GeV}^2$. The dashed line corresponds to $\bar{d}$ and the solid line to $\bar{u}$.}
\label{parameters}
\end{figure*}

The optimization converges within physically reasonable bounds, yielding stable parameter evolution. The resulting fit quality is good, with reduced $\chi^2/\text{d.o.f.} = 0.392$ for $\bar{u}$ and $0.249$ for $\bar{d}$. Uncertainties are estimated using the symmetric Hessian covariance matrix, with a tolerance rescaling applied to match the CT18 90\% confidence level. The extracted parameters are listed in Table~\ref{tab:evolution_parameters}. The scale dependence of shape parameters is shown in Fig.~\ref{parameters}.

\begin{table}[htpb]
\caption{Extracted evolution coefficients for the $\bar{u}$ and $\bar{d}$ sea quarks, determined via a continuous global fit to CT18NNLO data~\cite{hou2021new} over the range $\mu^2 \in [1.0, 100.0]$~GeV$^2$. The quadratic contributions are implemented using a squared Padé-like approximant to ensure stable high-scale behavior.}
\label{tab:evolution_parameters}
\begin{ruledtabular}
\begin{tabular}{ccc}
Parameter & $\bar{u}$ & $\bar{d}$ \\
\colrule
$c_{a_1}$ & $-0.675 \pm 0.002$ & $-0.386 \pm 0.003$ \\
$c_{a_2}$ & $\phantom{-}1.817 \pm 0.031$ & $\phantom{-}5.963 \pm 0.045$ \\
$c_{b_1}$ & $\phantom{-}14.903 \pm 0.084$ & $\phantom{-}3.241 \pm 0.071$ \\
$c_{b_2}$ & $-5.161 \pm 0.097$ & $-14.933 \pm 0.342$ \\
\colrule
$d_{a_1}$ & $\phantom{-}1.435 \pm 0.008$ & $\phantom{-}0.480 \pm 0.010$ \\
$d_{a_2}$ & $-8.787 \pm 0.123$ & $-24.504 \pm 0.180$ \\
$d_{b_1}$ & $-62.806 \pm 0.336$ & $-14.921 \pm 0.285$ \\
$d_{b_2}$ & $\phantom{-}17.318 \pm 0.404$ & $\phantom{-}58.799 \pm 1.454$ \\
\colrule
$\delta_1$ & $\phantom{-}3.460 \pm 0.008$ & $\phantom{-}2.847 \pm 0.009$ \\
$\delta_2$ & $\phantom{-}2.328 \pm 0.005$ & $\phantom{-}2.241 \pm 0.006$ \\
\end{tabular}
\end{ruledtabular}
\end{table}

\begin{figure*}[htpb]
\centering
\includegraphics[width=0.8\textwidth]{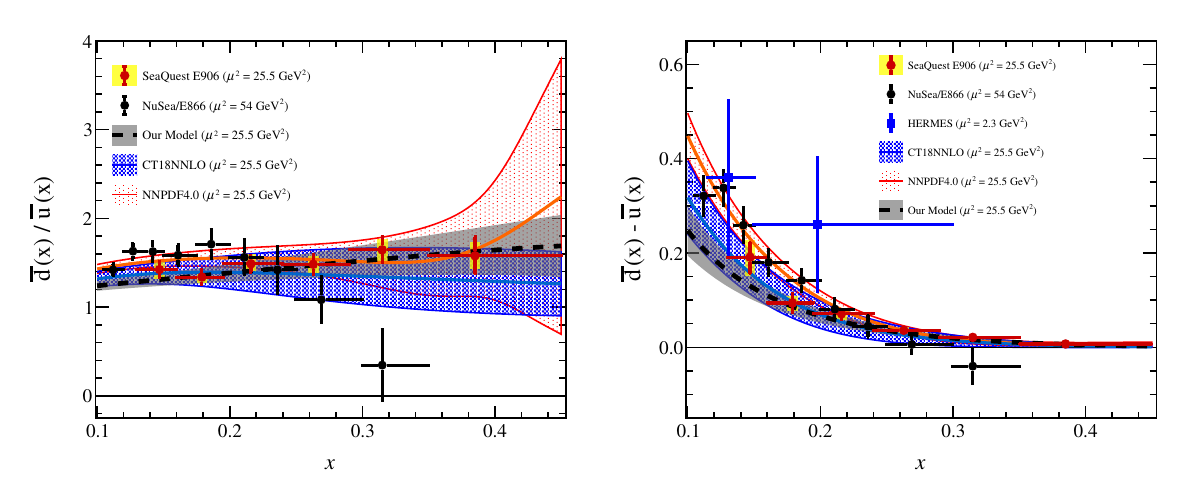}
\caption{The light sea-quark flavor asymmetry at $\mu^2 = 25.5$~GeV$^2$. Left: $\bar{d}/\bar{u}$. Right: $\bar{d}-\bar{u}$. Compared with CT18NNLO~\cite{hou2021new}, NNPDF4.0~\cite{ball2022path}, and data from SeaQuest~\cite{dove2023measurement}, NuSea~\cite{towell2001improved}, and HERMES~\cite{airapetian2004flavor}.}
\label{asymmetry}
\end{figure*}

Figure~\ref{asymmetry} shows the predicted flavor asymmetry at $\mu^2 = 25.5$~GeV$^2$. The model is consistent with the SeaQuest measurements~\cite{dove2023measurement} and modern global fits~\cite{hou2021new,ball2022path}, reproducing a sustained enhancement of $\bar{d}$ over $\bar{u}$ at intermediate-to-large $x$. This behavior contrasts with earlier NuSea results~\cite{towell2001improved}, which indicated a decrease at large $x$. The difference $\bar{d}-\bar{u}$ is also well reproduced, exhibiting the expected positive and smoothly decreasing trend.

To quantify this agreement, we evaluate truncated moments in the SeaQuest kinematic region ($0.13 \le x \le 0.45$):
\begin{align}
\int_{0.13}^{0.45} [\bar{d}(x) - \bar{u}(x)] , dx &= 0.0129 \pm 0.0056, \\
\int_{0.13}^{0.45} x[\bar{d}(x) - \bar{u}(x)] , dx &= 0.0025 \pm 0.0012.
\end{align}

These results are in very good agreement with CT18NNLO~\cite{hou2021new} and are consistent with experimental extractions from SeaQuest~\cite{dove2023measurement}. In contrast, other approaches such as NNPDF4.0~\cite{ball2022path}, the statistical model~\cite{basso2016drell}, and meson cloud models~\cite{alberg2019chiral,alberg2022pions} tend to yield larger values. This agreement highlights the ability of the present framework to capture the dominant nonperturbative dynamics of the nucleon sea.

\begin{table}[htbp]
\caption{Moments of the light sea-quark asymmetry evaluated at $\mu^2 = 25.5 \text{ GeV}^2$. The integrals are calculated over the SeaQuest kinematic range $x \in [0.13, 0.45]$.}
\label{moments}
\begin{ruledtabular}
\begin{tabular}{lcc}
Source & $ \int [\bar{d} - \bar{u}] dx$ & $\int x[\bar{d} - \bar{u}] dx$ \\
\colrule
SeaQuest  & $0.0159^{+0.0028+0.0028}_{-0.0030-0.0030}$ & $0.00318^{+0.00057+0.00055}_{-0.00062-0.00059}$ \\
Our Model & $0.0129 \pm 0.0056$ & $0.0025 \pm 0.0012$ \\
CT18 & $0.0129^{+0.0105}_{-0.0075}$ & $0.00241^{+0.00244}_{-0.00170}$ \\
NNPDF4.0 & $0.0208^{+0.0036}_{-0.0036}$ & $0.00414^{+0.00078}_{-0.00078}$ \\
Statistical & $0.0186$ & $0.00386$ \\
Meson Cloud & $0.0180$ & $0.00361$ \\
\end{tabular}
\end{ruledtabular}
\end{table}

We further examine the behavior of the flavor asymmetry at higher scales $\mu^2$ to assess the consistency of the model under evolution. Our light-front spectator framework evaluates the distributions $\bar{u}(x,\mu^2)$ and $\bar{d}(x,\mu^2)$ independently, which is well suited for describing the non-perturbative intrinsic sea in the moderate-to-large $x$ region.

At asymptotically small $x$ ($x \lesssim 0.01$), however, the dynamics are dominated by perturbative gluon splitting processes ($g \to q\bar{q}$), which generate an approximately flavor-symmetric sea. Consequently, the non-singlet difference $\bar{d}(x,\mu^2) - \bar{u}(x,\mu^2)$ is expected to be suppressed in this limit. Since the present model does not explicitly incorporate these singlet-driven small-$x$ dynamics, the independent parametric evolution of $\bar{u}$ and $\bar{d}$ can lead to residual non-singlet contributions in this region that lie outside the intrinsic sea component described by the model. Accordingly, all integrated flavor asymmetries are evaluated in the restricted region $x > 0.01$, where the intrinsic contribution dominates and the framework remains theoretically reliable.

The integrated moments of the asymmetry in the range $0.01 \leq x \leq 0.1$ are summarized in Table~\ref{momentssmall}.

\begin{table}[htbp]
\caption{Moments of the light sea-quark asymmetry evaluated at $\mu^2 = 200, 500,$ and $1000~\text{GeV}^2$ in the range $0.01 \leq x \leq 0.1$.}
\label{momentssmall}
\begin{ruledtabular}
\begin{tabular}{lcc}
Source & $\int [\bar{d} - \bar{u}]\, dx$ & $\int x[\bar{d} - \bar{u}]\, dx$ \\
\colrule
Our Model (200~GeV$^2$) & $0.06361 \pm 0.0230$ & $0.00251 \pm 0.0006$ \\
CT18 (200~GeV$^2$)      & $0.05585 \pm 0.0201$ & $0.00249 \pm 0.0006$ \\
\colrule
Our Model (500~GeV$^2$) & $0.05805 \pm 0.0279$ & $0.00243 \pm 0.0007$ \\
CT18 (500~GeV$^2$)      & $0.05581 \pm 0.0193$ & $0.00246 \pm 0.0006$ \\
\colrule
Our Model (1000~GeV$^2$)& $0.05048 \pm 0.0315$ & $0.00233 \pm 0.0008$ \\
CT18 (1000~GeV$^2$)     & $0.05572 \pm 0.0188$ & $0.00243 \pm 0.0006$ \\
\end{tabular}
\end{ruledtabular}
\end{table}

As seen from the table, the model exhibits good agreement with the CT18NNLO global fits across all considered scales. In particular, the momentum-weighted moment $\int x(\bar{d}-\bar{u})\,dx$ shows excellent stability and closely follows the CT18NNLO values, indicating that the dominant momentum contributions in the moderate-$x$ region are well captured by the evolved parametric form. The unweighted moment $\int (\bar{d}-\bar{u})\,dx$ shows a mild scale dependence but remains consistent within uncertainties, reflecting its residual sensitivity to the lower-$x$ boundary of the integration region. The gradual broadening of the model uncertainty at higher scales reflects the extrapolation of the fitted parameters beyond the primary fitting domain, while the behavior of the CT18NNLO uncertainty band is consistent with the expected smoothing of non-singlet distributions under perturbative QCD evolution.

\begin{figure*}[htpb]
\centering
\includegraphics[width=1.0\textwidth]{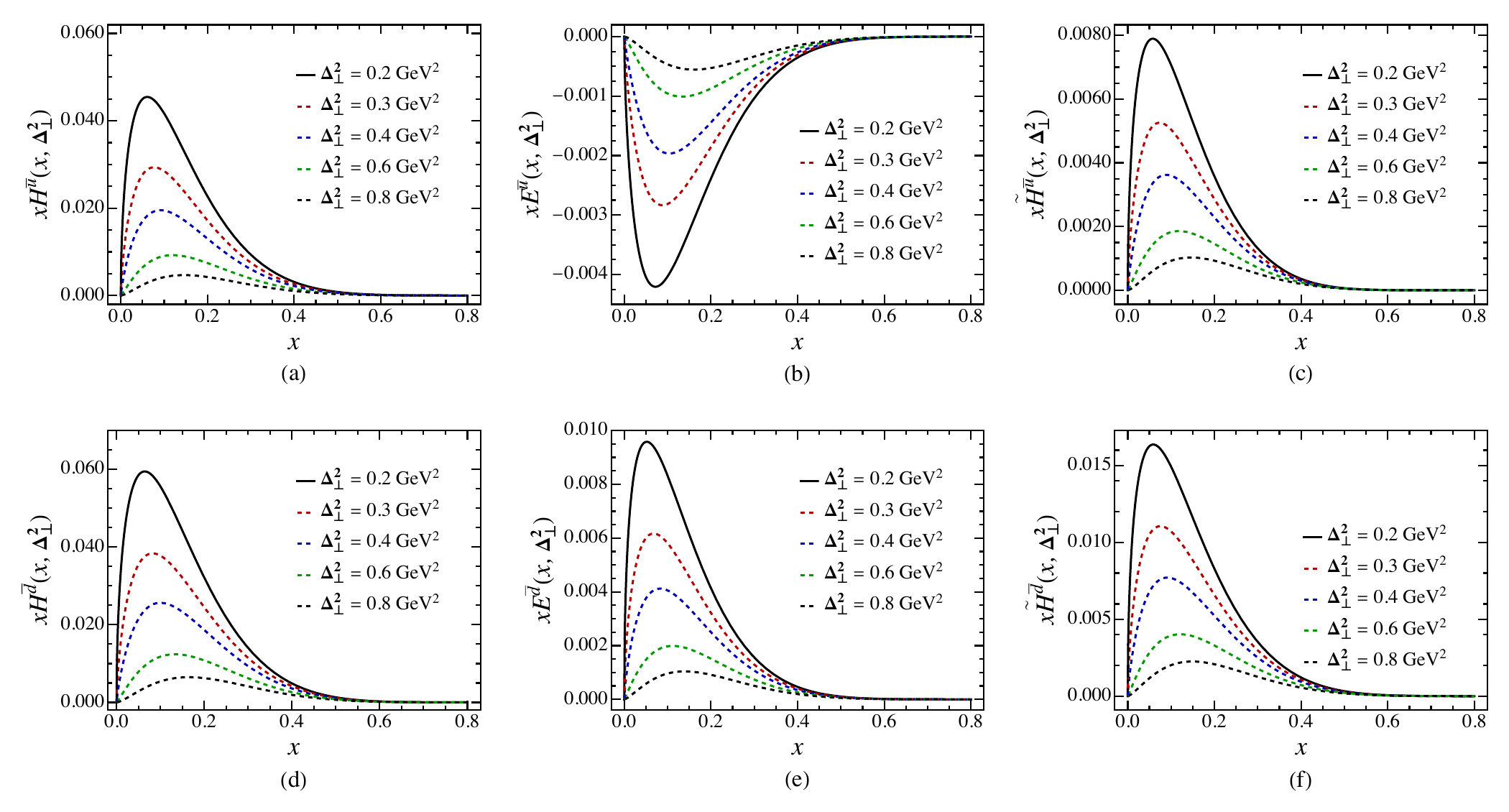}
\caption{The chiral even generalized parton distributions, $xH^{\bar{q}}(x,\boldsymbol{\Delta}_\perp^2)$, $xE^{\bar{q}}(x,\boldsymbol{\Delta}_\perp^2)$ and  $x\tilde{H}^{\bar{q}}(x,\boldsymbol{\Delta}_\perp^2)$, evaluated at zero skewness ($\xi = 0$) as functions of $x$ for various fixed values of $\boldsymbol{\Delta}_\perp^2$.}
\label{fig:gpds}
\end{figure*}

\section{Chiral-Even Generalized Parton Distributions}
\label{Sec:V}

At leading twist, Generalized Parton Distributions (GPDs) naturally extend our familiar unpolarized and helicity parton distributions into the off-forward regime. We define the chiral-even GPDs as the off-forward matrix elements of the light-front correlation functions for the vector and axial-vector currents~\cite{diehl2003generalized,diehl2001generalized}:

\begin{align}
&\frac{1}{2}\int\frac{dz^-}{2\pi}e^{ixP^+z^-} \langle p', \lambda' | \bar{\psi}_q(-z/2) \gamma^+ \psi_q(z/2) | p, \lambda \rangle \Big|_{\substack{z^+=0 \\ \mathbf{z}_\perp=0}} \notag \\
&\hspace{1cm} = \frac{1}{2P^+} \bar{u}(p', \lambda') \left[ H^q \gamma^+ + E^q \frac{i}{2M} \sigma^{+\alpha}\Delta_\alpha \right] u(p, \lambda), \\
&\frac{1}{2}\int\frac{dz^-}{2\pi}e^{ixP^+z^-} \langle p', \lambda' | \bar{\psi}_q(-z/2) \gamma^+ \gamma_5 \psi_q(z/2) | p, \lambda \rangle \Big|_{\substack{z^+=0 \\ \mathbf{z}_\perp=0}} \notag \\
&\hspace{1cm} = \frac{1}{2P^+} \bar{u}(p', \lambda') \left[ \tilde{H}^q \gamma^+\gamma_5 + \tilde{E}^q \frac{\gamma_5 \Delta^+}{2M} \right] u(p, \lambda).
\end{align}
where $p$ ($p'$) and $\lambda$ ($\lambda'$) represent the initial (final) momentum and helicity of the proton, and $M$ is the proton mass. Here, $H^q$ and $E^q$ are the unpolarized GPDs, while $\tilde{H}^q$ and $\tilde{E}^q$ are the helicity-dependent (axial) GPDs. 

Working in the symmetric frame, we define the standard kinematic variables: the average momentum $P^\mu = \frac{1}{2}(p+p')^\mu$, the momentum transfer $\Delta^\mu = p'^\mu - p^\mu$, the skewness $\xi = -\Delta^+ / 2P^+$, and the squared transfer $t = \Delta^2$. By choosing the light-front gauge ($A^+ = 0$), the gauge link between the quark fields reduces to unity. This allows us to map the GPDs directly to the helicity-dependent matrix elements of the proton and active quark:
\begin{equation}
A_{\lambda' \mu', \lambda \mu} = \int \frac{dz^-}{2\pi} e^{i x P^+ z^-} \langle p', \lambda' | \mathcal{O}_{\mu' \mu}(z) | p, \lambda \rangle \Big|_{z^+=0, \mathbf{z}_\perp=0}.
\end{equation}
The operators corresponding to definite quark helicities $\mu$ and $\mu'$ are given by $\mathcal{O}_{\pm,\pm} = \frac{1}{4} \bar{\psi} \gamma^+ (1 \pm \gamma_5) \psi$. Due to parity invariance, these amplitudes obey the symmetry relation $A_{-\lambda' -\mu', -\lambda -\mu} = (-1)^{\lambda'-\lambda} A_{\lambda' \mu', \lambda \mu}^*$. 

If we choose a reference frame where the momenta $\mathbf{p}$ and $\mathbf{p}'$ lie in the $x$-$z$ plane, we can isolate the individual chiral-even GPDs using specific combinations of these helicity amplitudes:
\begin{align}
H^q &= \frac{1}{\sqrt{1-\xi^2}} T_1^q - \frac{2M\xi^2}{\sqrt{t_0-t}(1-\xi^2)} T_3^q, \\
E^q &= -\frac{2M}{\epsilon\sqrt{t_0-t}} T_3^q, \\
\tilde{H}^q &= \frac{1}{\sqrt{1-\xi^2}} T_2^q + \frac{2M\xi}{\sqrt{t_0-t}(1-\xi^2)} T_4^q, \\
\tilde{E}^q &= \frac{2M}{\epsilon\xi\sqrt{t_0-t}} T_4^q.
\end{align}
Here, the matrix elements $T_i^q$ are constructed from the helicity basis as:
\begin{align}
T_1^q &= A_{++,++} + A_{-+,-+}, \\
T_2^q &= A_{++,++} - A_{-+,-+}, \\
T_3^q &= A_{++,-+} - A_{-+,++}, \\
T_4^q &= A_{++,-+} + A_{-+,++}.
\end{align}
The kinematic boundaries require a minimum momentum transfer $-t_0 = 4M^2\xi^2 / (1-\xi^2)$ for a given skewness $\xi$. The phase factor is defined as $\epsilon = \text{sgn}(D^1)$, where $D^1$ is the $x$-component of $D^\alpha = P^+ \Delta^\alpha - \Delta^+ P^\alpha$ (with $D^1 = 0$ exactly at $t=t_0$). At zero skewness condition only three GPDs are non zero~\cite{Choudhary:2023unw}.

The antiquark distributions are defined via this conjugate correlation function. Its relationship to the quark-quark correlation function is given by~\citep{kumano2021twist}:

\begin{equation}
F_{\Lambda' \Lambda}^{C[\Gamma]}(x, \xi, t) =
\begin{cases}
  \begin{aligned}
    &-F_{\Lambda' \Lambda}^{[\Gamma]}(-x, \xi, t) \\
    &\quad \text{for } \Gamma = \gamma^j, \gamma^-, i\sigma^{ij}\gamma_5, i\sigma^{j-}\gamma_5
  \end{aligned} \\[3ex]
  \begin{aligned}
    &+F_{\Lambda' \Lambda}^{[\Gamma]}(-x, \xi, t) \\
    &\quad \text{for } \Gamma = \mathbf{1}, \gamma^j \gamma_5.
  \end{aligned}
\end{cases}
\end{equation}

For the purposes of this study, we evaluate the system at zero skewness ($\xi = 0$), where the momentum transfer is purely transverse ($t = -\boldsymbol{\Delta}_\perp^2$). Note that while antiquark GPDs strictly reside in the $-x$ region, we evaluate and plot them against positive $x$ for clarity and ease of comparison. Evaluating the overlap representation within our spectator framework yields the following analytical expressions for the sea-antiquark GPDs:

\begin{widetext}
\begin{align}
H^{\bar{q}}(x,\boldsymbol{\Delta}_\perp^2) &= 
\left[ C_S^2 N_S^{(\bar{q})2} + C_V^2\left(\frac{1}{3}N_0^{(\bar{q})2} + \frac{2}{3}N_1^{(\bar{q})2}\right) \right] \nonumber \\ 
&\times \bigg[ \frac{1}{\delta^{\bar{q}}} x^{2a_1^{\bar{q}}} (1-x)^{2b_1^{\bar{q}}+1} 
+\frac{1}{M^2 \delta^{\bar{q}}} x^{2a_2^{\bar{q}}-2} (1-x)^{2b_2^{\bar{q}}+3} \left( \frac{\kappa^2}{\delta^{\bar{q}} \ln(1/x)} - \frac{\boldsymbol{\Delta}_\perp^2}{4} \right) \bigg]\exp\left(- \frac{\boldsymbol{\Delta}_\perp^2 \delta^{\bar{q}} \ln(1/x)}{4\kappa^2} \right),\\
E^{\bar{q}}(x, \boldsymbol{\Delta}_\perp^2)&=
2 \left( C_S^2 N_S^{(\bar{q})2} - \frac{1}{3} C_V^2 N_0^{(\bar{q})2} \right) 
\frac{(1-x)^2}{\delta^{\bar{q}}} x^{a_1^{\bar{q}} + a_2^{\bar{q}} - 1} (1-x)^{b_1^{\bar{q}} + b_2^{\bar{q}}}
 \times\exp\left(- \frac{\boldsymbol{\Delta}_\perp^2 \delta^{\bar{q}} \ln(1/x)}{4\kappa^2} \right),\\
\tilde{H}^{\bar{q}}(x, \boldsymbol{\Delta}_\perp^2)
&=
\left[ C_S^2 N_S^{(\bar{q})2} + C_V^2\left(\frac{1}{3}N_0^{(\bar{q})2} - \frac{2}{3}N_1^{(\bar{q})2}\right) \right] \nonumber \\
&\times \Bigg[ \frac{1}{\delta^{\bar{q}}} x^{2a_1^{\bar{q}}} (1-x)^{2b_1^{\bar{q}}+1} 
- \frac{1}{M^2 \delta^{\bar{q}}} x^{2a_2^{\bar{q}}-2} (1-x)^{2b_2^{\bar{q}}+3} \left( \frac{\kappa^2}{\delta^{\bar{q}} \ln(1/x)} - \frac{\boldsymbol{\Delta}_\perp^2}{4} \right) \Bigg] 
\exp\left(- \frac{\boldsymbol{\Delta}_\perp^2 \delta^{\bar{q}} \ln(1/x)}{4\kappa^2} \right).
\end{align}
\end{widetext}

We present the computed sea quark Generalized Parton Distributions (GPDs) as a function of the longitudinal momentum fraction $x$ for various values of the transverse momentum transfer squared, $\Delta_{\perp}^{2}$. The distributions $xH^{\overline{q}}(x,\Delta_{\perp}^{2})$, $xE^{\overline{q}}(x,\Delta_{\perp}^{2})$, and $x\tilde{H}^{\overline{q}}(x,\Delta_{\perp}^{2})$ are illustrated in figure~\ref{fig:gpds}. The top panels (a, b, and c) correspond to the anti-up ($\overline{u}$) quark, while the bottom panels (d, e, and f) represent the anti-down ($\bar{d}$) quark distributions.

We observe that all the GPDs exhibit pronounced peaks in the low-$x$ region (typically around $x \approx 0.05$ to $0.15$) and fall off rapidly, approaching zero for $x > 0.6$. Furthermore, we notice a consistent trend across all distributions: the magnitudes of the GPDs systematically decrease as the momentum transfer $\Delta_{\perp}^{2}$ increases from 0.2~GeV$^2$ to 0.8~GeV$^2$. The maximum amplitude for each distribution corresponds to the lowest momentum transfer ($\Delta_{\perp}^{2} = 0.2$~GeV$^2$).

Comparing the unpolarized GPDs, both $xH^{\overline{u}}$ and $xH^{\overline{d}}$ are positive definite. However, the distribution for the anti-down quark exhibits a larger magnitude, with $xH^{\overline{d}}$ reaching a peak value of approximately 0.06, compared to $xH^{\overline{u}}$ which peaks near 0.045. A similar magnitude enhancement for the $\overline{d}$ quark is observed in the helicity-dependent GPDs, where $x\tilde{H}^{\overline{d}}$ peaks around 0.016 while $x\tilde{H}^{\overline{u}}$ peaks around 0.006. Both $x\tilde{H}^{\overline{u}}$ and $x\tilde{H}^{\overline{d}}$ maintain positive distributions.

Notably, a distinct flavor asymmetry is evident in the magnetic GPDs, $xE^{\overline{q}}$. The distribution $xE^{\overline{u}}(x,\Delta_{\perp}^{2})$ shows distinctly different behavior from the others by taking purely negative values, with its largest negative peak reaching approximately $-0.004$. In contrast, $xE^{\overline{d}}(x,\Delta_{\perp}^{2})$ is strictly positive, featuring a maximum peak around $0.010$. This opposing sign between the $\overline{u}$ and $\overline{d}$ quarks in the $E$ distribution highlights a significant flavor dependence, which is particularly relevant when evaluating the orbital angular momentum (OAM) contributions of individual sea quark flavors to the total nucleon spin.

Our qualitative results for the sea-quark GPDs broadly align with existing theoretical and phenomenological studies. The positive definite nature of $xH^{\bar{q}}$ for both flavors is consistent with nonlocal chiral effective theory~\cite{he2022generalized} and perturbative QCD (pQCD) frameworks~\cite{kriesten2022parametrization}. Likewise, the monotonic decrease of the GPDs with increasing momentum transfer matches both nonlocal chiral models~\cite{he2022generalized} and global fits~\cite{fazio2017physics}. Crucially, the opposing signs we observe in the magnetic GPDs ($xE^{\bar{u}} < 0$ and $xE^{\bar{d}} > 0$) agree with nonlocal chiral theory. However, this contradicts certain pQCD parameterizations~\cite{kriesten2022parametrization} that predict the exact opposite. 

\subsection{Total angular momentum}

To quantify the contribution of sea antiquarks to the nucleon spin, we evaluate the Ji sum rule~\cite{ji1997gauge}. This relation provides a gauge-invariant decomposition of the proton's angular momentum in terms of the second moments of the unpolarized and Pauli GPDs. In the forward limit ($t \to 0$, $\xi \to 0$), the total angular momentum carried by a given antiquark flavor is given by
\begin{equation}
    J^{\bar{q}} = \frac{1}{2} \int_{0}^{1} dx \, x \left[ H^{\bar{q}}(x, 0, 0) + E^{\bar{q}}(x, 0, 0) \right].
\end{equation}

\begin{figure}[htpb]
    \centering
    \includegraphics[width=0.8\linewidth]{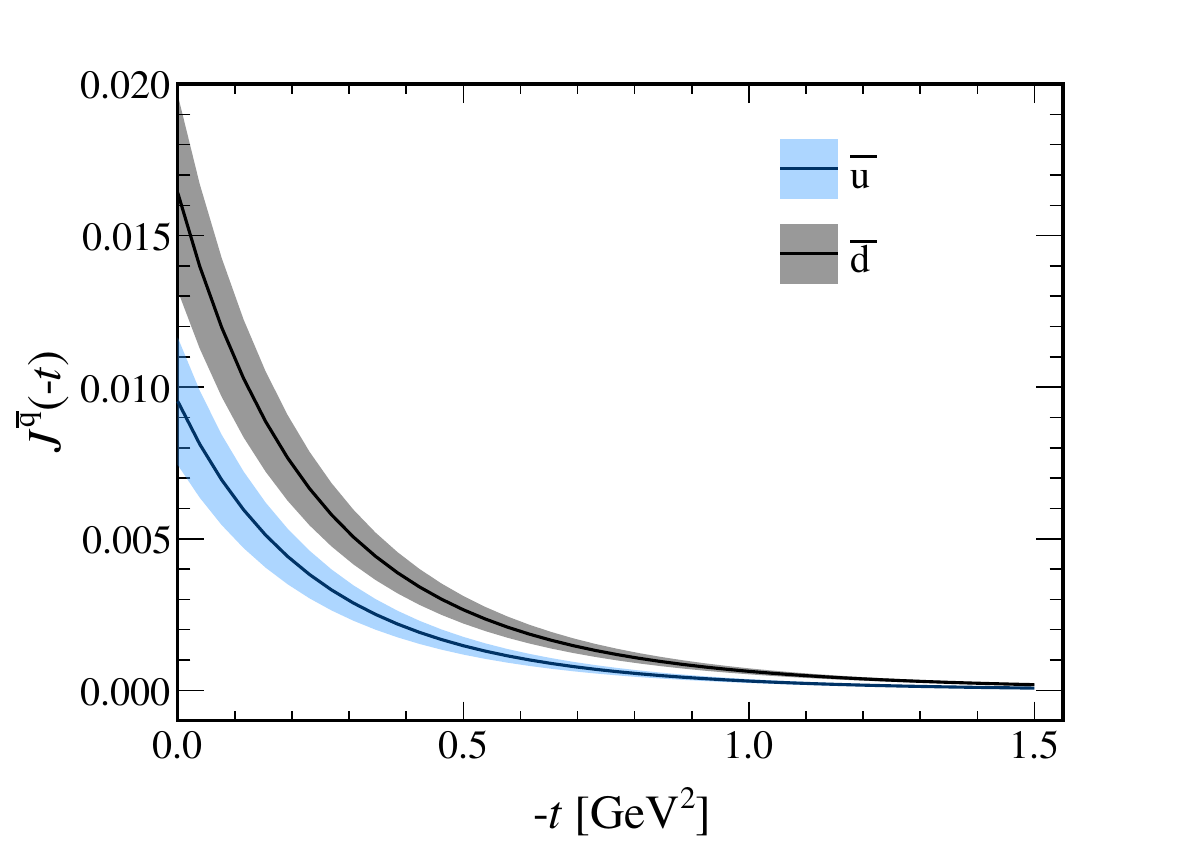}
    \caption{Generalized angular momentum form factor $J^{\bar{q}}(t)$ as a function of momentum transfer $-t$. The black curve corresponds to $\bar{d}$, while the blue curve represents $\bar{u}$.}
    \label{Jq}
\end{figure}

To illustrate the off-forward structure, we evaluate the generalized angular momentum form factor
\begin{equation}
    J^{\bar{q}}(t) = \frac{1}{2} \int_0^1 dx\, x \left[ H^{\bar{q}}(x,0,t) + E^{\bar{q}}(x,0,t) \right]
\end{equation}
and plot it as a function of $-t$ at the initial scale $\mu_0^2 = 1.0~\text{GeV}^2$ in Fig.~\ref{Jq}. The distributions exhibit a smooth decrease with increasing momentum transfer, consistent with the expected falloff of form factors.

Evaluating the Ji sum rule using the evolved parameters at $\mu^2 = 4.0$~GeV$^2$, we obtain
\begin{align}
J^{\bar{u}} &= 0.0122 \pm 0.0029, \\
J^{\bar{d}} &= 0.0179 \pm 0.0038.
\end{align}
These results indicate a clear flavor asymmetry, with $J^{\bar{d}} > J^{\bar{u}}$, showing that down antiquarks carry a larger fraction of the nucleon's angular momentum.

Extracting sea-quark angular momentum from experimental data remains challenging. Phenomenological fits, such as the Goloskokov-Kroll model~\cite{Diehl:2013xca}, are weakly constrained in the sea sector and often assume $E^{\bar{q}} \approx 0$, leading to $J^{\bar{q}} \approx 0$ with large uncertainties. In contrast, extractions based on the Sivers function~\cite{Bacchetta:2011gx} yield
$J^{\bar{u}} = 0.015 \pm 0.003$ and $J^{\bar{d}} = 0.022 \pm 0.005$ at $\mu^2 = 4.0$~GeV$^2$. Our results are consistent with these values within uncertainties and reproduce the observed flavor asymmetry.

Lattice QCD calculations provide a complementary constraint through disconnected contributions. Results from the $\chi$QCD~\cite{Deka:2013zha} and ETMC~\cite{Alexandrou:2017oeh} collaborations yield small positive values for the sea angular momentum, with $\chi$QCD reporting $J^{\bar{u}}_{\text{sea}} = J^{\bar{d}}_{\text{sea}} \approx 0.018 \pm 0.004$. Due to isospin symmetry in these calculations, flavor asymmetries are not resolved. The flavor-averaged value from our model, $\langle J^{\bar{q}} \rangle \approx 0.0151$, is in good agreement with these lattice estimates.

Within our scalar-vector spectator framework, the flavor asymmetry arises naturally. The enhancement $J^{\bar{d}} > J^{\bar{u}}$ is consistent with expectations from meson-cloud models~\cite{Thomas:2008bd}, where $p \to n + \pi^+$ fluctuations generate orbital angular momentum predominantly carried by down antiquarks. This provides a natural explanation for the connection between the observed $\bar{d} > \bar{u}$ asymmetry and the distribution of angular momentum in the nucleon.

\section{CONCLUSION}
\label{Conclusion}

In this work, we developed an effective light-front spectator model to investigate the nonperturbative sea-quark structure of the nucleon. By treating the proton as an active sea antiquark paired with a composite scalar-vector spectator, we successfully incorporated the necessary spin-flavor symmetries and spin-orbit correlations. We parametrized the spatial light-front wave functions using a soft-wall AdS/QCD profile~\cite{maji2016light} and constrained our initial parameters via a simultaneous global fit to CT18NNLO PDFs~\cite{hou2021new} and Bacchetta-Radici\cite{Bacchetta:2011gx} phenomenological extractions at the initial scale $\mu_0^2 = 1.0~\text{GeV}^2$.

To evaluate high-energy phenomena without directly integrating the coupled DGLAP equations, we implemented a continuous, parameter-driven scale evolution. At the SeaQuest kinematic scale of $\mu^2 = 25.5~\text{GeV}^2$, our model explicitly predicts a sustained excess of $\bar{d}$ over $\bar{u}$ at high $x$. This confirms the $\bar{d}$ enhancement observed by the E906 experiment~\cite{dove2023measurement} and demonstrates that our spectator framework naturally captures the nonperturbative dynamics driving the light sea-quark flavor asymmetry.

Furthermore, we calculated the leading chiral-even generalized parton distributions $xH^{\bar{q}}(x,\boldsymbol{\Delta}_\perp^2)$, $xE^{\bar{q}}(x,\boldsymbol{\Delta}_\perp^2)$ and  $x\tilde{H}^{\bar{q}}(x,\boldsymbol{\Delta}_\perp^2)$ in momentum space at zero skewness. By evaluating the Ji sum rule with these distributions, we extracted the total angular momentum of the individual sea quarks. We observed a significant flavor asymmetry, finding that $J^{\bar{d}} >J^{\bar{u}}$. Notably, our flavor-averaged total angular momentum is in excellent agreement with Sivers transverse-momentum distribution results~\cite{Bacchetta:2011gx}. These results offer a physically intuitive and robust description of the sea-quark GPDs, providing a valuable framework for interpreting future high-precision experimental data.

\bibliography{ref}

\end{document}